\documentclass[aps,
pra,
twocolumn,
superscriptaddress,
nolongbibliography,
nofootinbib,
]{revtex4-2}
\usepackage[utf8]{inputenc}
\usepackage{amsmath}
\usepackage{amssymb}
\usepackage{graphicx}
\usepackage{color,soul}
\usepackage{siunitx}
\usepackage{dsfont}
\usepackage{bm}
\usepackage{braket}
\usepackage[
	hidelinks,
	colorlinks=true,
	linkcolor=black,
	citecolor=blue,
	urlcolor=black,
]{hyperref}\usepackage[english]{babel}
\usepackage{lineno} \usepackage{etoolbox}

\usepackage[english,nomargin,inline,marginclue,draft]{fixme}
\fxusetheme{colorsig}
\FXRegisterAuthor{db}{adb}{\color{blue}DB} \FXRegisterAuthor{th}{ath}{\color{green}TH} \FXRegisterAuthor{tc}{atc}{TC}
\makeatletter
\renewcommand*\FXLayoutInline[3]{{\@fxuseface{inline}\ignorespaces{\color{fx#1}[#3: #2]}}}
\makeatother
\def\nobreakbefore{\relax\ifvmode\else
    \ifhmode
      \ifdim\lastskip > 0pt\relax
        \unskip\nobreakspace
      \else \nobreakspace
      \fi
    \fi
  \fi
}
\let\oldcite\cite
\renewcommand\cite{\nobreakbefore\oldcite}

\newcommand{\ee}{\ensuremath{\mathrm{e}}}
\newcommand{\ii}{\ensuremath{\mathrm{i}}}
\newcommand{\figref}[1]{Fig.~\ref{#1}}
\newcommand{\Eqref}[1]{Eq.~\eqref{#1}}
\newcommand{\secref}[1]{Sec.~\ref{#1}}
\newcommand{\tabref}[1]{Tab.~\ref{#1}}
\newcommand{\appref}[1]{Appendix~\ref{#1}}
\DeclareSIUnit\Gauss{G}
\DeclareSIUnit\rad{rad}
\DeclareSIUnit\mrad{mrad}
\DeclareSIUnit\Er{\text{$\mathrm{E}^{\mathrm{R}}$}}
\DeclareSIUnit\Eri{\text{$\mathrm{E}_{\ensuremath{i}}^{\mathrm{R}}$}}
\DeclareSIUnit\Erx{\text{$\mathrm{E}_{\ensuremath{x}}^{\mathrm{R}}$}}
\DeclareSIUnit\Ery{\text{$\mathrm{E}_{\ensuremath{y}}^{\mathrm{R}}$}}
\DeclareSIUnit\Erxl{\text{$\mathrm{E}_{\ensuremath{x},\mathrm{L}}^{\mathrm{R}}$}}
\DeclareSIUnit\ab{\text{$a_\mathrm{B}$}}
\usepackage{booktabs}

\makeatletter
\def\maketitle{
\@author@finish
\title@column\titleblock@produce
\suppressfloats[t]}
\makeatother

\makeatletter\begin{document}

\newcommand{\MPQ}{Max-Planck-Institut f\"{u}r Quantenoptik, 85748 Garching, Germany}
\newcommand{\MCQST}{Munich Center for Quantum Science and Technology, 80799 Munich, Germany}
\newcommand{\LMU}{Fakult\"{a}t f\"{u}r Physik, Ludwig-Maximilians-Universit\"{a}t, 80799 Munich, Germany}
\newcommand{\LCF}{Laboratoire Charles Fabry, Institut d'Optique Graduate School, CNRS, Universit\'e Paris-Saclay, 91127 Palaiseau, France}

\title{Optical superlattice for engineering Hubbard couplings in quantum simulation}

\author{Thomas~Chalopin}\email[Electronic address: ]{thomas.chalopin@institutoptique.fr}\affiliation{\MPQ}\affiliation{\MCQST}\affiliation{\LCF}

\author{Petar~Bojovi\'c}\affiliation{\MPQ}\affiliation{\MCQST}

\author{Dominik~Bourgund}\affiliation{\MPQ}\affiliation{\MCQST}

\author{Si~Wang}\affiliation{\MPQ}\affiliation{\MCQST}

\author{Titus~Franz}\affiliation{\MPQ}\affiliation{\MCQST}

\author{Immanuel~Bloch}\affiliation{\MPQ}\affiliation{\MCQST}\affiliation{\LMU}

\author{Timon~Hilker}\affiliation{\MPQ}\affiliation{\MCQST}

\begin{abstract}

Quantum simulations of Hubbard models with ultracold atoms rely on the exceptional control of coherent motion provided by optical lattices.
Here we demonstrate enhanced tunability using an optical superlattice in a fermionic quantum gas microscope.
With our phase-stable bichromatic design, we achieve a precise control of tunneling and tilt throughout the lattice, as evidenced by long-lived coherent double-well oscillations and next-nearest-neighbor quantum walks in a staggered configuration.
We furthermore present correlated quantum walks of two particles initiated through a resonant pair-breaking mechanism.
Finally, we engineer tunable spin couplings through local offsets and create a spin ladder with ferromagnetic and antiferromagnetic couplings along the rungs and legs, respectively.
Our work underscores the high potential of optical superlattices for engineering, simulating, and detecting strongly correlated many-body quantum states.

\end{abstract}

\maketitle

\paragraph*{Introduction ---}

Ultracold atoms confined in optical lattices have proven to be an exceptionally fruitful approach for exploring, understanding, and engineering quantum many-body phases \cite{bloch:2008}.
During recent years, significant progress has been made in quantum simulations of Hubbard models, especially with quantum gas microscopes \cite{gross:2021}, which resolve individual particles of a degenerate quantum gas at each lattice site \cite{bakr:2009,sherson:2010} and allow manipulation of the quantum system with local control \cite{weitenberg:2011, choi:2016, islam:2015, ji:2021, sompet:2022, hirthe:2023,young:2024}.
Lattices beyond the simple square geometry allow to engineer different band structures \cite{becker:2010, tarruell:2012, jo:2012, duca:2015, cooper:2019}, explore strongly-correlated magnetic phases \cite{gall:2021}, simulate artificial magnetic fields \cite{aidelsburger:2011}, study topological charge pumping \cite{lohse:2016, nakajima:2016, lohse:2018}, prepare out-of-equilibrium states \cite{schreiber:2015}, implement quantum gates \cite{anderlini:2007, trotzky:2008, yang:2020, zhang:2023}, induce frustration \cite{yang:2021, xu:2023, prichard:2024} and improve detection \cite{impertro:2023}.

An optical superlattice consists of two (or more) superimposed optical lattices with commensurate lattice constants.
In the context of quantum simulation, effective state preparation, simulation, and readout methods include folded lattices \cite{sebby-strabley:2006}, single-wavelength lattices with phase stabilization \cite{tarruell:2012}, and bichromatic lattices with \cite{folling:2007, schreiber:2015, dai:2016, gall:2021} or without phase stabilization \cite{koepsell:2020, li:2021a}.
A major challenge of such experimental platforms lies in achieving minimal phase noise for enhanced quantum coherence while maintaining a high degree of tunability for state engineering and dynamics.
State-of-the-art platforms have reported superlattice phase stability as low as \SI{10}{\mrad} \cite{li:2021a, impertro:2023}, and even sub-\si{\mrad} in tunable honeycomb lattices \cite{kosch:2022}.

Here, we demonstrate enhanced state preparation, dynamics, and quantum simulation of the Fermi-Hubbard model with a phase-stable bichromatic superlattice.
We show single-atom control within isolated double-wells (DWs), showcasing substantial coherence time through Rabi oscillations.
We furthermore investigate strongly correlated quantum walk dynamics in a one-dimensional staggered configuration, revealing direct control over nearest-neighbor (NN) and next-nearest-neighbor (NNN) tunnel couplings.
Finally, we engineer Fermi-Hubbard ladders with tunable couplings, where we demonstrate full control over spin superexchange --- including the inversion of its sign ---, resulting in intriguing hybrid ferromagnetic--antiferromagnetic systems.
\begin{figure*}[!t]
\centering
\includegraphics[scale = 1]{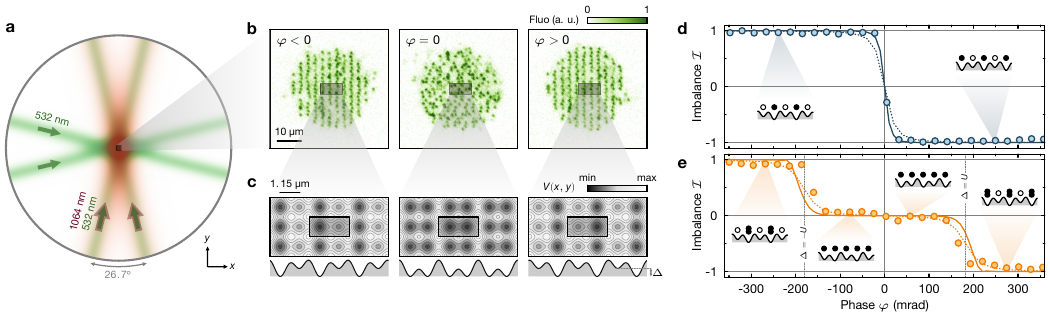}
\caption{
\textbf{Experimental setup.}
\textbf{a}, The superlattice is engineered by superimposing optical lattices generated from the interference of laser beams at \SI{532}{\nm} and \SI{1064}{\nm} under an angle.
In this work, only the short lattice (\SI{532}{\nm}) is used in the $y$ direction.
\textbf{b}, Exemplary site-resolved fluorescence pictures of atoms for different relative phases $\varphi$.
\textbf{c}, Lattice potential for the phases shown in \textbf{b}, with the unit cell marked by the black rectangle.
\textbf{d, e}, Imbalance $\mathcal{I}$ of the system in the case of one atom per unit cell (\textbf{d}) or two atoms per unit cell (\textbf{e}).
The error bars in this figure, as in all other figures, are derived from a bootstrap procedure and are smaller than the data points.
The lines are obtained from the ground state of a two-site Fermi-Hubbard model with parameters $t$, $U$, and $\Delta$ (see main text) obtained from \emph{ab-initio} Wannier function calculations with (solid lines) and without (dashed lines) accounting for the finite ramp time of our detection procedure \cite{supp}. }
\label{fig:fig1}
\end{figure*}

\paragraph*{Description of the system ---}
Our quantum simulator realizes the spin-1/2 repulsive Fermi-Hubbard model using ultracold $^{6}$Li loaded in two-dimensional optical superlattices generated from the interference of laser beams at \SI{532}{\nm} and \SI{1064}{\nm} under an angle of \SI{26.7}{\degree} (see \figref{fig:fig1}a).
Our bichromatic design is based on a single laser source at \SI{1064}{\nm} whose beam is split, with one part directly generating the long lattice and another part amplified and directed into a second harmonic generation (SHG) setup to generate laser light at \SI{532}{\nm} for the short lattice.
Our design ensures an intrinsic stability of the relative phase $\varphi$ between the two lattices and was described in detail in our previous work \cite{koepsell:2020}.
A delay line in one of the two interfering beams, in combination with dynamic frequency shifting of the \SI{532}{\nm} beam, allows tuning of the relative phase $\varphi$ and control of the local potential seen by the atoms (\appref{sec:appA}).
Our apparatus is equipped with superlattices in the $x$ and $y$ directions, but in this work, we only use them along $x$.

The total potential $V(x, y)$ imprinted to the atoms takes the form
\begin{align}
	V(x, y) =\ & V_{x}\cos^2[\pi x/a_x + \varphi] + V_{x,\mathrm{L}}\sin^2[\pi x/(2a_x)]\notag\\
	& + V_y\sin^2[\pi y/a_y],
	\label{eq:eq1}
\end{align}
where $a_x = \SI{1.15(1)}{\um}$, $a_y = \SI{1.11(1)}{\um}$ are the lattice constants, $V_x$, $V_y$ are the short lattice depths along $x$ and $y$, and $V_{x,\mathrm{L}}$ is the long lattice depth along $x$.
In the following, the lattice depths are expressed in units of their respective recoil energy $\mathrm{E}_{i}^{\mathrm{R}} = h^2/8ma_i^2$ ($i = x, y$) for the short lattices and $\mathrm{E}_{x,\mathrm{L}}^{\mathrm{R}} = \mathrm{E}_{x}^{\mathrm{R}}/4$ for the long lattice, with $h$ the Planck's constant and $m$ the mass of a single atom.

The unit cell of the system contains two lattice sites, and the potential shape within a unit cell varies continuously upon varying the phase $\varphi$. Staggered potentials with an energy offset on the even (odd) sites for $\varphi < 0$ ($\varphi > 0$) can thus be engineered, while a balanced configuration is obtained for $\varphi = 0$ (\figref{fig:fig1}c).
The atomic ensemble is well described by the single-band Fermi-Hubbard model, with Hamiltonian
\begin{align}
    \hat{H} &= \sum_{\braket{ij},\sigma}\left[-t_{ij}\hat{c}^\dag_{i,\sigma}\hat{c}_{j,\sigma} + \mathrm{h.c.}\right] \notag \\
    &\quad + U\sum_{i}\hat{n}_{i,\uparrow}\hat{n}_{i,\downarrow} + \sum_{i,\sigma}\Delta_i\hat{n}_{i,\sigma},
    \label{eq:eq2}
\end{align}
where $\braket{ij}$ are NN sites, $\sigma = \uparrow, \downarrow$ is the spin state, $\hat{c}^\dag_{i,\sigma}$ is the fermionic creation operator for spin $\sigma$ at site $i$, $t_{ij}$ is the tunnel coupling between sites $i$ and $j$, $\hat{n}_{i,\sigma}=\hat{c}^\dag_{i,\sigma}\hat{c}_{i,\sigma}$ is the atom number operator at site $i$ for spin $\sigma$, $U > 0$ is the on-site interaction energy, and $\Delta_i$ is a site-dependent energy offset.
In this work, $t_{ij} = t_y$ if $i, j$ are NN along $y$, and $t_{ij} = t_x^{(1)}$ ($t_x^{(2)}$), for $i, j$ NN along $x$ within (between) a unit cell(s).
The tunneling energies $t_{ij}$ and offsets $\Delta_i$ are controlled by the lattice potential from \Eqref{eq:eq1}, and the interaction energy between the two spin states, encoded by the lowest two hyperfine states of $^{6}\mathrm{Li}$, is set using the broad Feshbach resonance around \SI{830}{\Gauss}.

The superlattice configuration directly impacts the atomic density distribution, which we measure with single-site resolution by performing fluorescence imaging (\figref{fig:fig1}b).
In particular, we use it to calibrate and characterize the lattice phase control of our apparatus by loading a balanced mixture of both spin states in the superlattice at different phases $\varphi$.
The lattice depths are chosen in order to engineer a system of quasi-isolated DWs, with tunneling amplitude in the balanced configuration $t_x^{(1)}/h = \SI{510(73)}{\Hz}$ and $t_x^{(2)}/h = \SI{13(1)}{\Hz}$ (\appref{sec:appC}).

We measure the normalized imbalance $\mathcal{I} = (\braket{\hat{n}_{\mathrm{o}}} - \braket{\hat{n}_{\mathrm{e}}})/(\braket{\hat{n}_{\mathrm{o}}} + \braket{\hat{n}_{\mathrm{e}}})$ (\figref{fig:fig1}d,e), where $\braket{\hat{n}_{\mathrm{e(o)}}}$ is the average atomic density on even (odd) sites.
When considering DWs populated with a single atom (\figref{fig:fig1}d), the symmetric phase $\varphi = 0$ is identified as the phase for which a balanced system ($\mathcal{I} = 0$) is engineered.
As the phase changes, only the lowest well is populated in each unit cell, resulting in a strong shift in the imbalance towards $|\mathcal{I}| \approx 1$ away from the symmetric configuration.
In particular, we measure an average imbalance $|\bar{\mathcal{I}}| = \num{0.985(2)}$ in the large tilt regime $\SI{50}{\mrad} \leq |\varphi| \leq \SI{300}{\mrad}$ ($\num{4.3(6)} \leq |\Delta/t_x^{(1)}| \leq \num{23(4)}$).

\begin{figure}[!t]
\centering
\includegraphics[scale=1]{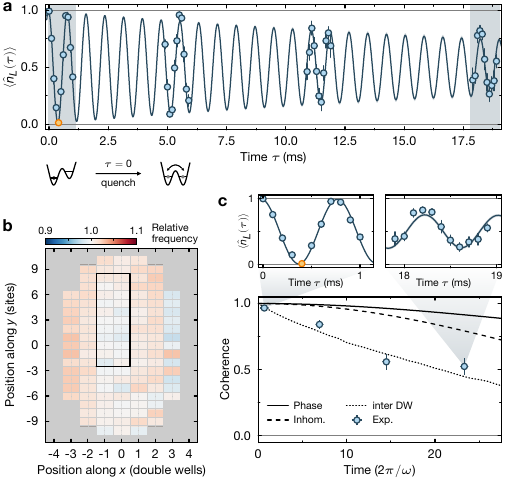}
\caption{
\textbf{Double-well oscillations.}
\textbf{a}, Population evolution $\braket{\hat{n}_{\mathrm{L}}(\tau)}$ in the left site of the DWs as a function of time $\tau$ after the quench ($\tau = 0$) from an imbalanced situation ($\varphi < 0$) to a symmetric double well ($\varphi = 0$).
The orange point is used to evaluate our experimental $\pi$-pulse fidelity (see text).
\textbf{b}, Local DW oscillation frequency.
The black rectangle indicates the 22 DWs considered in \textbf{a}.
\textbf{c}, Coherence of the measured oscillations (blue points), compared to different decoherence models (see text).
}
\label{fig:fig2}
\end{figure}

When considering DWs populated with two atoms, a fully imbalanced configuration $|\mathcal{I}| \approx 1$ can only be reached for a tilt that is sufficient to overcome the interaction energy, $|\Delta_i| > U$.
Thus, we observe a balanced configuration $\mathcal{I} = 0$ over a region $|\varphi| \lesssim \varphi_{\mathrm{c}}$, with $\varphi_{\mathrm{c}} \approx \SI{170}{\mrad}$ in our configuration where $U/h \approx \SI{7.7}{\kHz}$, while a fully imbalanced configuration is reached for $|\varphi| \gtrsim \varphi_c$.
In \figref{fig:fig1}d and e, the data is well matched by a ground-state calculation of a two-site Fermi-Hubbard model without free parameters (solid lines).
\paragraph*{Double-well oscillations ---}
We highlight the phase stability of our system by conducting DW oscillations.
The system is initialized away from the symmetric phase at $\varphi = \SI{-400}{\mrad}$, resulting in a fully imbalanced configuration, and the average density is set to be about one atom per double well.
The relative phase is then quenched to the balanced configuration $\varphi = 0$ at time $\tau = 0$, leading to an oscillation of the populations of the two sites in the double well (see \figref{fig:fig2}a).
In our data analysis, we post-select on DWs containing exactly one detected atom.

The evolution is well captured by a resonant two-level oscillation with dephasing $\braket{\hat{n}_{\mathrm{L}}(\tau)}=[1+\cos(\omega\tau)\ee^{-\tau/\tau_\mathrm{d}}]/2$.
The extracted oscillation frequency $\omega=2\pi\times\SI{1.261(1)}{\kHz}$ is in good agreement with the expected frequency $\omega_{\mathrm{th}} = 2t_x^{(1)}\hbar^{-1} = 2\pi \times \SI{1.36(20)}{\kHz}$ calculated from the lattice depths (\appref{sec:appC}).
Our apparatus furthermore allows probing DW oscillations locally, revealing spatial inhomogeneities in the oscillation frequency (\figref{fig:fig2}b), which can be attributed to the inhomogeneities of the underlying lattice potential \cite{supp}.
The decay time $\tau_\mathrm{d} = \SI{27(3)}{\ms} = \num{33(4)} \times 2\pi/\omega$ is found to be consistent with residual tunneling between neighboring DWs (\figref{fig:fig2}c), due to the finite depths of our lattices.
Moreover, we find that this residual tunneling is the dominant source of decoherence, compared to spatial inhomogeneities across the system and phase fluctuations (\appref{sec:appA}, see also \cite{supp}).

The two-level system defined by the DW potential can be interpreted as an orbital qubit \cite{impertro:2023}, the quality of the DW oscillations directly assessing the achievable fidelity of single-qubit gate operations.
In particular, we directly measure a $\pi$-pulse fidelity $P_{\pi}^{\text{exp}}=1-\braket{\hat{n}_{\mathrm{L}} (\tau = \pi/\omega)} = 0.988^{+0.007}_{-0.009}$ (orange point in \figref{fig:fig2}a,c) by averaging over 22 DWs in the center of our system (black rectangle in \figref{fig:fig2}b).
Using the fit to the data, which takes into account dephasing and post-selection (see above), we find $P_{\pi} = \num{0.991(1)}$.
When taking into account the detection-induced errors associated with the motion and loss of particles during imaging, this fidelity increases to $P_{\pi} = \num{0.993(1)}$ \cite{supp}.
We note that our primary source of decoherence is associated with inter-DW coupling, such that we expect the fidelity to improve if larger long lattice depths are achieved --- in practice, we estimate $P_\pi > 0.999$ for reasonable depths $V_{x,\mathrm{L}} \gtrsim \SI{70}{\Erxl}$.

\paragraph*{Quantum walks ---}

To illustrate how our superlattice can be exploited to tune tunnel couplings, we study larger-scale dynamics of our system \emph{via} quantum walks in one dimension.
These walks are carried out in two lattice configurations: in a standard lattice potential ($V_{x,\mathrm{L}} = 0$) and in a staggered potential ($V_{x,\mathrm{L}} >0, \varphi = \pi/2$, equal tunneling $t_x^{(1)} = t_x^{(2)} = t$).
We use our Digital Micromirror Device (DMD) to populate a single column of atoms along $y$ in a frozen configuration, and the dynamics along $x$ is initiated at $\tau = 0$ by abruptly quenching the short and long lattices along $x$ to lower depths (\appref{sec:appC}).
The average atomic density $\braket{\hat{n}_i(\tau)}$ is then reconstructed as a function of space and time.

\begin{figure*}[!t]
\centering
\includegraphics[scale = 1]{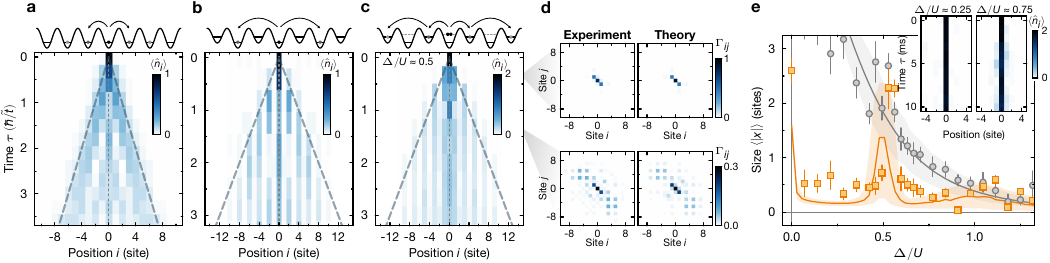}
\caption{
\textbf{Quantum walks on a superlattice.}
\textbf{a, b}, Symmetrized density distributions $\braket{\hat{n}_i(\tau)}$ for a state localized at $i = 0$ at $\tau = 0$, in the case of a standard lattice \textbf{(a)} and a staggered superlattice at $\varphi = \pi/2$ \textbf{(b)}.
In \textbf{a}, the particle delocalizes on a timescale given by the NN tunneling energy $\tilde{t} = t = h\times\SI{96(11)}{\Hz}$.
In \textbf{b}, direct NN coupling is suppressed, but NNN processes remain resonant with an effective coupling $\tilde{t} = t^2/\Delta + t' = h\times\SI{64(11)}{\Hz}$ (see main text).
\textbf{c}, A superlattice initialized with a doublon features a resonant pair-breaking process upon tilting at $\Delta = U/2$, where the doublon is coupled to a delocalized state of two singly occupied neighbors.
At longer times, the distribution behaves similarly to \textbf{b}, with $\tilde{t} = t^2/(-\Delta) + t' = h\times\SI{52(9)}{\Hz}$.
The dashed lines in \textbf{a}, \textbf{b}, \textbf{c} indicate the expected scaling behavior $x \sim 2\tilde{t}\tau/\hbar$ of the quantum walk expansion. 
\textbf{d}, Correlation maps $\Gamma_{ij} = \braket{\hat{c}_i^\dag\hat{c}_j^\dag\hat{c}_j\hat{c}_i}$ of the experimental data and from numerical simulations.
\textbf{e}, Spatial extension of the density distribution after a fixed time $\tau = \SI{4}{\ms}$ ($\tau t/\hbar \approx 8$, with $t$ the bare tunnel coupling) as a function of the staggering offset $\Delta/U$, and for singly and doubly occupied initial states (gray circles and orange squares, respectively).
Solid lines are obtained from numerical simulations, the shading indicating the uncertainty (one std) resulting from the uncertainty of the Fermi-Hubbard parameters.
(Inset) Examples of time evolutions of the density distribution where the resonant condition is not met showcase the absence of delocalization.
}
\label{fig:fig3}
\end{figure*}

We first consider the dynamics of a single atom in a standard lattice (\figref{fig:fig3}a).
We post-select the experimental data on rows populated with one atom ($\sum_i\hat{n}_i = 1$), and we recover the expected dynamics from quantum walks \cite{supp}.
The time axis is here given in units of tunneling times $(\tilde{t}/\hbar)^{-1}$, where $\tilde{t}$ is an effective tunneling amplitude.
For a standard lattice, it is simply the NN tunneling energy $\tilde{t} = t = h \times \SI{96(11)}{\Hz}$ (\appref{sec:appC}).

We then consider the same experiment in a staggered potential (\figref{fig:fig3}b), with $t/h = \SI{320(25)}{\Hz}$, and an energy offset at neighboring sites $\Delta/h = \SI{1.36(7)}{\kHz}$, which almost completely suppresses NN tunneling.
NNN sites remain degenerate in energy, resulting in a non-zero effective coupling $\tilde{t}=t'+t^2/\Delta$.
Here, $t'/h = -\SI{12(2)}{\Hz}$ is the direct tunneling process between the NNN sites, and $t^2/\Delta=h\times\SI{76(13)}{\Hz}$ is a perturbative coupling that depends on the virtual population of the intermediate (NN) site \cite{supp}.
Such a staggered configuration results in a quantum walk whose dynamics takes place only on the sublattice that contains the initial state.
Even in these slower dynamics, we observe coherent evolution up to almost \num{10} lattice sites before the expansion slows down due to large-scale inhomogeneities of our lattices inherent to the Gaussian envelope of the laser beams \cite{supp}.

Interaction effects are revealed by post-selecting data with two atoms per row, ($\sum_i\hat{n}_i = 2$), and total spin $0$, effectively considering dynamics with the initial state being a doubly occupied site (a doublon).
Such an initial state corresponds to a repulsively bound state \cite{folling:2007, winkler:2006, preiss:2015}.
In the case of a staggered lattice considered here, the motion of the doublon is generally fully suppressed, as an NNN coupling would be a 4$^{\mathrm{th}}$ order process in the tunneling energy \cite{supp}.
However, a resonant pair-breaking process can be found by tuning the NN energy offset to be close to half the interaction energy, $\Delta = U/2$. Here, the initial state, corresponding to a doublon at position $i = 0$, is degenerate with the state corresponding to one atom in each of the neighboring sites $i = \pm 1$, allowing a breaking of the bound pair.
We show in \figref{fig:fig3}c a quantum walk in this configuration, depicting population in sites $i = \pm 1$ in the early dynamics, as a consequence of this pair-breaking process.
Subsequent dynamics appear to be similar to the one in \figref{fig:fig3}b, with non-zero population on the sublattice that contains sites $i = \pm 1$.

Despite being spatially separated at longer times, the atoms forming the initial bound state remain correlated.
The correlation map $\Gamma_{ij}=\braket{\hat{c}_i^\dag\hat{c}_j^\dag\hat{c}_j\hat{c}_i}=\braket{\hat{n}_i\hat{n}_j}-\braket{\hat{n}_i}\delta_{ij}$, with $\delta_{ij}$ the Kronecker delta, shown in \figref{fig:fig3}d for two evolution times, $\tau\tilde{t}/\hbar\approx1$ and $\tau\tilde{t}/\hbar\approx\num{3.5}$, reveals long-range non-zero correlations between two atoms distant up to \num{12} sites for the latter.
The agreement with theory, taking into account lattice inhomogeneities \cite{supp}, is excellent, indicating that the quantum coherence of the evolution is maintained over long time.

The difference between singly and doubly occupied initial states is striking when considering how much the atomic distribution has spread after a fixed time.
We show in \figref{fig:fig3}e the extension of the atomic density distribution $\braket{|x|}=\sum_i p_i(\tau)|x_i|$, with $p_i(\tau) = \braket{\hat n_i(\tau)}/\sum_i\braket{\hat n_i(\tau)}$, for $\tau = \SI{4}{\ms}$ ($t\tau/\hbar \approx 8$), as a function of the staggering energy offset $\Delta/U$.
In the case of a singly occupied initial state (gray circles), we measure a localization of the distribution when increasing $\Delta$, following the reduction of the effective tunneling energy $\tilde{t}$.
In the case of a doubly occupied initial state (orange squares), we rather observe a very sharp resonance around $\Delta/U \approx 0.5$, corresponding to the pair-breaking situation of \figref{fig:fig3}c, while the distribution remains localized for most other staggering offsets (examples are given in the inset).
The situation at $\Delta = 0$ corresponds to the quantum walk of a bound state on a standard lattice, with an effective tunneling rate following a scaling similar to that for singlons in a staggered lattice.
Closer to $\Delta = U$, a weak delocalization is expected due to the direct (non-perturbative) coupling between the initial bound pair with energy $U$ and a split configuration with a particle at $i = 0$ and another one at $i = \pm 1$.
\paragraph*{Engineering ferromagnetic couplings ---}
While a staggered configuration allows suppressing tunneling, magnetic correlations are predicted to remain, except in the resonant configuration $\Delta \approx U$ for which a non-vanishing doublon population is expected \cite{duan:2003, trotzky:2008}.
Away from this resonant condition, NN spin interactions are expected to be described by a perturbative superexchange coupling $J$ which changes sign for $\Delta > U$,
\begin{equation}
	J(\Delta) = \frac{4t^2/U}{1 - (\Delta/U)^2}.
	\label{eq:eq3}
\end{equation}

\begin{figure}[!t]
\centering
\includegraphics[scale = 1]{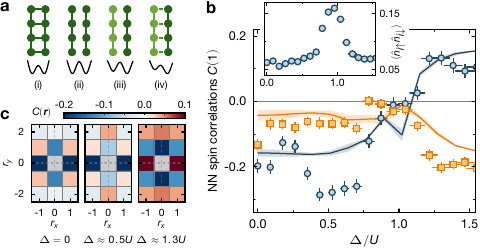}
\caption{
\textbf{Tuning the sign of the superexchange coupling.}
\textbf{a}, The preparation sequence consists in 4 steps: (i) loading a balanced ladder configuration, (ii) decoupling the legs of the ladder, (iii) applying an offset $\Delta$ on one leg and (iv) re-coupling the legs.
\textbf{b}, NN spin correlations along $x$ (blue) and $y$ (orange) as a function of $\Delta$.
The solid lines are numerical results obtained from finite-temperature exact diagonalization (see text).
(Inset) Doublon density as a function of $\Delta$.
\textbf{c}, Example of spin correlation maps, illustrating the sign inversion of the spin coupling along $x$.
}
\label{fig:fig4}
\end{figure}

We explore this effect by loading an ensemble of ladders of length $L = 11$ with $\braket{\hat{n}} \approx 0.93$ atoms per site, with Hubbard parameters $t_x/h = \SI{342(49)}{\Hz}$ ($t_y/h = \SI{163(20)}{\Hz}$) along the rungs (legs) and $U/h = \SI{4.96(12)}{\kHz}$ at $\varphi=0$ (\appref{sec:appC}).
The loading procedure (\figref{fig:fig4}a) relies on first freezing the system along $x$ [steps (i) -- (ii)] before applying the tilt with a variable superlattice phase $\varphi$ [steps (iii) -- (iv)], in order to prevent atoms from exclusively populating the lower leg in the regime of strong tilts $\Delta>U$.
This procedure is very similar to the one used in our recent work \cite{hirthe:2023, bourgund:2023} to engineer mixed-dimensional (mixD) systems.

We evaluate the normalized spin-spin correlations,
\begin{equation}
	C(\bm{r}) = \frac{1}{\mathcal{N}_{\bm{r}}}\sum_{\substack{i,j \\ i-j=\bm{r}}}\frac{\langle\hat{S}_i^{z}\hat{S}_j^{z}\rangle - \langle\hat{S}_i^{z}\rangle\langle\hat{S}_{j}^{z}\rangle}{\sigma(\hat{S}_i^z)\sigma(\hat{S}_j^z)},
\end{equation}
with $\sigma^2(\hat{S}_j^z) = \braket{(\hat{S}_i^z)^2} - \braket{\hat{S}_i^z}^2$ the on-site spin fluctuations, and $\mathcal{N}_{\bm{r}}$ a normalization constant counting the number of pairs of sites of size $\bm{r}$.
We show in \figref{fig:fig4}b the NN spin correlations, $C(|\bm{r}| = 1)$, as a function of tilt $\Delta$.
The inset indicates the doublon density $\braket{\hat n_{i\uparrow}\hat n_{i\downarrow}}$ in the system, showing that an excess number of doubly occupied sites are created only in the vicinity of $\Delta \approx U$.

At $\Delta = 0$, the NN spin correlations are close to $-0.2$, with an asymmetry between the $x$ and $y$ directions, due to the anisotropy of the superexchange coupling ($J_x/J_y \approx 4$).
As $\Delta$ increases, the superexchange coupling increases along the $x$-direction, while it remains unchanged along $y$.
We observe an increase in the magnitude of the spin correlations along $x$ and a slight decrease along $y$, as a consequence of the redistribution of the correlations to favor the $x$ direction.
The highest correlations are reached around $\Delta/U \approx 0.5$, which corresponds to a regime where the rung tunneling is suppressed, while the doublon density remains low.
As $\Delta$ approaches $U$, we observe vanishing spin correlations in all directions, concomitant to the peak in the doublon density.

When tilting above the interaction energy, \emph{i.e.} $\Delta > U$, the superexchange coupling along the $x$ direction changes sign, as expected from \Eqref{eq:eq3}.
In this regime, we observe positive spin-spin correlations along $x$, signaling ferromagnetic coupling along the rungs of the ladders, while antiferromagnetic ordering remains along $y$.

We compare the experimental data with finite-temperature exact diagonalization performed on a Fermi-Hubbard system of size $2\times4$ at half-filling ($\braket{\hat{n}} = 1$, insets).
The simulation assumes an initial thermal state at $k_{\mathrm{B}}T/t_x = 0.2$, and the full time-dependent Hamiltonian evolution is carried out using the \SI{50}{\ms} ramp used in the experiment \cite{supp}.
Our experimental results are qualitatively well reproduced by numerical simulations (solid lines in \figref{fig:fig4}b), from which one can clearly identify the different regimes realized in the experiment: enhanced antiferromagnetic rung correlations for $0 < \Delta <U$, vanishing correlations around $\Delta \approx U$, and ferromagnetic rung correlations for $\Delta > U$.
Our simulations also suggest that the prepared states remain closed to the equilibrium states of an effective mixD $t-J$ model without a drastic increase in temperature \cite{bourgund:2023, supp}.

We show in \figref{fig:fig4}c the symmetrized spin correlation maps $C(\bm{r})$, in three different regimes, $\Delta/U = 0$, $\Delta/U \approx 0.5$, and $\Delta/U \approx 1.3$.
For $\Delta < U$, the characteristic checkerboard antiferromagnetic pattern is visible, with strong rung singlets along $x$ in the large $J_x$ regime.
For $\Delta > U$, the change in sign of the superexchange coupling yields a dramatic change in the spin correlation map, which no longer features a checkerboard pattern.
The correlations are ferromagnetic along $x$, while antiferromagnetism is enhanced along $y$ due to the lack of competition in the spin ordering, as expected from the hybrid coupling engineered here.
\paragraph*{Conclusion ---}
The different experiments conducted in this work, ranging from single particle to many-body physics, illustrate how superlattices can be used in the context of quantum simulation of the Fermi-Hubbard model using ultracold atoms.
Our results yield significant prospects for future quantum simulations and quantum computations.
The DW oscillations, for instance, can be interpreted as a quantum gate operation of an orbital qubit, and can be readily extended to realize two-qubit gates, specifically in the form of collisional gates \cite{jaksch:1999, calarco:2000}, opening exciting prospects for quantum computation with fermionic particles \cite{bravyi:2002, naldesi:2023, sun:2023a, gonzalez-cuadra:2023}.
Additionally, the pair-breaking mechanism and subsequent coherent two-atom quantum walk realizes a novel regime of correlated quantum evolution compared to previous work \cite{folling:2007, preiss:2015, kwan:2023}.
It finds applications in the simulation of lattice gauge theories \cite{yang:2020a, zhou:2022, halimeh:2023} and can play a role in engineered quantum many-body systems \cite{lebrat:2024a}.

More generally, we have demonstrated how a superlattice can be used to control and modify the coupling strengths of the Hubbard Hamiltonian.
The fine tunability of tunneling and superexchange terms highlights the potential for the simulation of a broader range of real or theoretically interesting artificial materials.
A recent example is nickelate materials, for which mixed-dimensionality \cite{grusdt:2018, bohrdt:2022a, schlomer:2023b} appears to be crucial for the emergence of a superconducting phase at around \SI{80}{\K} \cite{sun:2023, qu:2024}.
MixD Fermi-Hubbard models enable the observation of charge-ordered states, like hole pairs or stripes, at experimentally accessible temperatures \cite{hirthe:2023, bourgund:2023}.
The ability to engineer ferromagnetic correlations in a many-body system at equilibrium goes beyond previous work in DWs \cite{trotzky:2008, honda:2023} and opens avenues for exploring symmetry-protected topological states \cite{sompet:2022}, engineering spin-1 systems, and simulating exotic Hubbard models \cite{putikka:1992, japaridze:2000}.
Combined with extended state readout schemes granted by superlattices \cite{atala:2014,schweizer:2016,kaufman:2016,impertro:2023}, such a toolbox considerably extends the capabilities of our quantum simulator, allowing us to experimentally realize settings that have not yet been accessed in traditional solid-state experiments.

\begin{acknowledgments}
We thank David Clément for insightful discussions and careful reading of the manuscript.
This work was supported by the Max Planck Society (MPG), the Horizon Europe program HORIZON-CL4-2022 QUANTUM-02-SGA (project 101113690, PASQuanS2.1), the German Federal Ministry of Education and Research (BMBF grant agreement 13N15890, FermiQP), and Germany's Excellence Strategy (EXC-2111-390814868).
T.C. acknowledges funding from the Alexander v. Humboldt Foundation.
\end{acknowledgments}

\appendix

\section{Superlattice phase control and stability}
\label{sec:appA}
The superlattice potential results from the superposition of two shallow-angle optical lattices produced from laser light at \SI{532}{\nm} and \SI{1064}{\nm} for the short and long lattice respectively (\figref{fig:fig1}a).
Each lattice is generated by the interference of two laser beams --- the two \emph{arms} of the lattice ---, with a difference in path length of about $L \approx \SI{40}{\cm}$.
A change in the frequency of the laser field translates into a phase shift of the interference pattern as a consequence of this delay line.
 When the short lattice frequency is changed from $f = f_0 = c/(\SI{532}{\nm})$ to $f = f_0 + \Delta f$, with $c$ the speed of light, the phase of the lattice potential changes by $\varphi = L\pi\Delta f/c$ (see \Eqref{eq:eq1}) .
Experimentally, the frequency shift is induced by Acousto-Optic Deflectors (AOD), in a similar manner as in \cite{koepsell:2020}.
In our setup, this method allows tuning the relative phase $\varphi$ by $\sim1.3\pi$, which is enough to go from a fully balanced configuration to a fully staggered configuration.

\begin{figure}[!t]
\centering
\includegraphics[scale = 1]{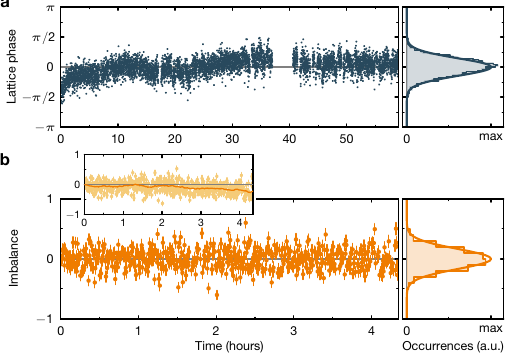}
\caption{
\textbf{Absolute and relative phase stability}
\textbf{a}, Absolute phase measured over the course of more than two days.
\textbf{b}, Relative phase stability, inferred from the repetitive measurement of the average imbalance in the system over a few hours.
Shot-to-shot fluctuations are extracted from the subtraction of long-term drifts (inset), evaluated by a rolling average over a period of 20 minutes (inset, solid orange line).
}
\label{fig:figS1}
\end{figure}

In the design process of the optical lattices, special care has been put into thermally and mechanically isolating the optical setup for splitting and recombining the lattice arms.
The optical elements, in particular, are glued on a glass material with low thermal expansion coefficient and placed in an evacuated box.
The independent propagation in air of each arm is limited to about \SI{10}{\cm}.
Additional details can be found in \cite{bourgund:2024}.

We evaluate the absolute phase of the setup through repetitive, long-term measurements of the same atomic distribution.
The absolute lattice phase, \emph{i.e.} the absolute position of the lattice grid with respect to the acquisition camera, is inferred for each experimental realization.
The results are given in \figref{fig:figS1}a, and illustrate the outstanding stability of our setup, featuring sub lattice-site fluctuations over a period of more than two days.

\begin{table*}[!ht]
\centering
\caption{
\textbf{Summary of experimental parameters.}
A range of parameters indicates that the parameter was varied during data acquisition.
A blank cell indicates irrelevant parameters: the interaction strength for single-particle experiments and the relative phase when the long lattice is off.
}

\setlength{\tabcolsep}{.26cm}
\renewcommand{\arraystretch}{1.2}
\begin{tabular}{l l l l l l l l l
}
\toprule
\textbf{Param.} & \textbf{Unit} & \textbf{\figref{fig:fig1}} & \textbf{\figref{fig:fig2}} & \textbf{\figref{fig:fig3}a} & \textbf{\figref{fig:fig3}b} & \textbf{\figref{fig:fig3}c} & \textbf{\figref{fig:fig3}e} & \textbf{\figref{fig:fig4}} \\
\midrule \midrule
$V_x$ & $\si{\Erx}$ & \num{11.0(5)} & \num{10.0(5)} & \num{11.0(5)} & \num{6.0(3)} & \num{6.0(3)} & \num{6.0(3)} & \num{11.0(5)} \\
$V_{x,\mathrm{L}}$ & $\si{\Erxl}$ & \num{31(1)} & \num{31(1)} & \num{0.0} & \num{1.00(5)} & \num{1.9(1)} & [\num{0}, \num{4}] & \num{23(1)} \\
$V_{y}$ & $\si{\Ery}$ & \num{40(2)} & \num{40(2)} & \num{40(2)} & \num{40(2)} & \num{40(2)} & \num{40(2)} & \num{9.0(5)} \\
$\varphi$ & \si{\mrad} & [\num{-400}, \num{400}] & \num{0} & --- & $\pi/2$ & $\pi/2$ & $\pi/2$ & [\num{0}, \num{240}]\\ 
$a_\mathrm{s}$ & $a_\mathrm{B}$ & \num{1293} & --- & --- & --- & \num{973} & \num{973} & \num{1293} \\
\midrule \midrule
$t_x^{(1)}/h$ & $\si{\Hz}$ & \num{510(73)}, \num{604(97)}\footnote{\label{footnoteFig1}Values for $\varphi = \num{0}$ and $\varphi = \SI{400}{\mrad}$ resp.} & \num{675(101)} & \num{97(13)} & \num{320(25)} & \num{323(25)} & \num{334(27)}\footnote{\label{footnoteFig3}Values for $V_{x,\mathrm{L}} = \SI{4.0(2)}{\Erxl}$.} & \num{342(49)}, \num{349(51)}\footnote{\label{footnoteFig4}Values for $\varphi = \num{0}$ and $\varphi = \SI{240}{\mrad}$ resp.}\\
$t_x^{(2)}/h$ & $\si{\Hz}$ & \num{13(1)}, \num{19(2)}\footref{footnoteFig1} & \num{14(1)} & \num{96(11)} &  \num{320(25)} & \num{323(25)} & \num{334(27)}\footref{footnoteFig3} & \num{22(2)}, \num{24(2)}\footref{footnoteFig4}\\
$t_y/h$ & $\si{\Hz}$ & $<\num{1}$ & $<\num{1}$ & $<\num{1}$ & $<\num{1}$ & $<\num{1}$ & $<\num{1}$ & \num{163(20)}\\
$t'/h$ & $\si{\Hz}$ & $<\num{1}$ & $<\num{1}$ & $<\num{1}$ & \num{12(2)} & \num{14(2)} & \num{16(3)}\footref{footnoteFig3} & $<\num{1}$ \\
$U/h$ & $\si{\kHz}$ & \num{7.53(16)}, \num{7.84(15)}\footref{footnoteFig1} & --- & --- & --- & \num{4.83(1)} & \num{4.88(10)}\footref{footnoteFig3} & \num{4.96(12)}, \num{5.04(11)}\footref{footnoteFig4}\\
$\Delta/h$ & $\si{\kHz}$ & 0, \num{16.7(5)}\footref{footnoteFig1} & --- & --- & --- & \num{2.58(14)} & \num{5.43(27)}\footref{footnoteFig3} & \num{0}, \num{7.66(32)}\footref{footnoteFig4} \\
\bottomrule
\end{tabular}
\label{tab:tab1}
\end{table*} 
The relative phase stability is inferred by performing repeated imbalance measurements (see \figref{fig:fig1}d) in the balanced configuration ($\varphi = 0$).
The average imbalance $\mathcal{I}$ is evaluated for each repetition, and represented vs. time in \figref{fig:figS1}b.
Fluctuations of the imbalance around $\mathcal{I} = 0$ are attributed to shot-to-shot fluctuation of the relative superlattice phase.
In particular, we measure a standard deviation of the imbalance $\sqrt{\Delta \mathcal{I}^2} = 0.167$, corresponding to phase fluctuations $\sqrt{\Delta\varphi^{2}} \approx \SI{4.5}{\mrad}$ according to the relation between imbalance and phase experimentally measured in \figref{fig:fig1} (thus already accounting for the effects of the freezing ramps).
Note that these fluctuations are only slightly larger than what is expected from the shot noise $\sqrt{\Delta \mathcal{I}^2_{\mathrm{shot}}} = 1/\sqrt{N}$, associated with the finite number of double-wells considered here ($N = \num{66}$).
We also measure long-term drifts on timescales of several hours (inset).
Over the course of data acquisition, these drifts are regularly compensated, to ensure proper preparation of the system.

\section{Experimental sequences}
\label{sec:appC}

All our sequences start with a 2D degenerate Fermi gas with a balanced mixture of both spin states, loaded into a single fringe of a vertical lattice with lattice constant $a_z = \SI{3}{\um}$.
In-plane confinement is ensured by a DMD-shaped repulsive potential.
The atomic density is controlled by varying the total atom number and the size of the in-plane confinement.

The atoms are then loaded into optical lattices, with the configurations varying depending on the specific experiment being performed.
For the experiments presented in \figref{fig:fig2} and \figref{fig:fig3}, the experimental procedure involves some dynamics initiated by quenching the superlattice relative phase or depth to a specified value (see below for details).
For every sequence, the final step consists of freezing the atomic distribution by ramping up the short lattices to $V_i \approx \SI{43}{\Eri}$ ($i = x, y$) in \SI{1.5}{\ms}, before performing a spin-resolved fluorescence image of the system \cite{koepsell:2020}.

We give in \tabref{tab:tab1} the lattice parameters for each of the measurements presented in this work.
Additional information on the sequences are given below.
More details on the calculation of the Hubbard parameters, can be found in the SI \cite{supp}.

\paragraph{Identification of the balanced configuration (\figref{fig:fig1})}

Before ramping up the lattice potentials, the relative phase $\varphi$ is set to the value at which the measurement is performed.
$V_y$ and $V_{x,\mathrm{L}}$ are then ramped up in \SI{50}{\ms} to their final value (see \tabref{tab:tab1}), thus making a lattice with lattice constants $2a_x$ and $a_y$.
After \SI{5}{\ms} of holding, $V_x$ is ramped up to its final value in \SI{50}{\ms}, splitting each lattice site along the $x$-direction in two.
Freezing is performed after \SI{2}{\ms} of holding in the final configuration.
\paragraph{Double-well oscillations (\figref{fig:fig2}).}
The relative phase $\varphi$ is initialized to be far from the balanced configuration around \SI{-400}{\mrad}.
$V_{x,\mathrm{L}}$ is ramped up first in \SI{200}{\ms} to its final value (see \tabref{tab:tab1}), making 1D tubes separated by $2a_x$ along the $x$-direction.
After \SI{20}{\ms} of holding, both $V_{x}$ and $V_y$ are ramped up to their final values in \SI{200}{\ms}.
Such a procedure allows to load on average one atom per double-well, assuming that the overall density is well set, and the interaction strength is large enough.
The lattice depths are held at these values for \SI{20}{\ms} before freezing.
The double-well oscillations are triggered by quenching the relative phase $\varphi$ to the balanced configuration in \SI{45}{\us} at a time $\tau \in (\SI{0}{\ms}, \SI{20}{\ms}]$ before freezing. The freezing procedure, in this particular case, also involves quenching $\varphi$ back to the initial phase. The data point at $\tau = \SI{0}{\ms}$ is taken without quench.
\paragraph{Quantum walks (\figref{fig:fig3}).}
For each of the quantum walk experiments, a tailored DMD potential restricts the system size to a single row of lattice sites along $y$, which is identified as position $x = 0$.
$V_y$ and $V_{x,\mathrm{L}}$ are ramped up together in \SI{200}{\ms} to about \SI{40}{\Ery} and \SI{31}{\Erxl}, thus making a single 1D array of atoms.
Only then is the short lattice turned on, as $V_x$ is ramped up to \SI{40}{\Erx} in \SI{200}{\ms}.
This two-step procedure facilitates initial state preparation, as loading a single array of atoms in the long lattice is easier than directly loading the short lattice.
The relative phase is always initialized beforehand to the configuration with maximum contrast ($\varphi = \pi/2$), which ensures that all the population of a single long-lattice site is transferred to a single short-lattice site upon ramping up $V_x$.
The system is held frozen for \SI{9}{\ms}, before $V_{x,\mathrm{L}}$ is ramped down to its final value (see \tabref{tab:tab1}) in \SI{20}{\ms}.
The dynamics is triggered by quenching $V_x$ to its final value in \SI{1}{\ms}.
The end of the quench ramp marks the beginning of the evolution of duration $\tau$ before freezing, and the data at $\tau = \SI{0}{\ms}$ is acquired without quench.
\paragraph{Ferromagnetic ladders (\figref{fig:fig4}).}
The relative phase is initialized in the balanced configuration $\varphi = 0$.
Then, all the lattices are ramped up in \SI{250}{\ms}, going through configuration (i) and (ii) in \figref{fig:fig4}a.
The intermediate lattice depths after step (ii) are $V_x \approx \SI{30}{\Erx}$, $V_{x,\mathrm{L}} = \SI{23(1)}{\Erxl}$ and $V_y \approx \SI{7}{\Ery}$.
The tilt is applied by quenching the relative phase $\varphi$ to its final value in \SI{0.5}{\ms}.
The short lattices are then ramped to their final value (see \tabref{tab:tab1}) in \SI{45}{\ms}.
Freezing is performed after \SI{0.5}{\ms} in this final configuration.

\clearpage
\newpage
\makeatletter
\renewcommand{\thefigure}{S\@arabic\c@figure}
\renewcommand{\theequation}{S\arabic{equation}}
\makeatother
\setcounter{figure}{0}
\setcounter{table}{0}
\setcounter{section}{0}
\setcounter{equation}{0}
\appendix

\title{Supplemental Material: \\ Optical superlattice for engineering Hubbard couplings in quantum simulation}

\maketitle

\subsection{Gate fidelities}

The fidelities of the double-well oscillations presented in \figref{fig:fig2} are experimentally limited by different sources of dephasing, including amplitude inhomogeneities, phase fluctuations and coupling to neighboring double wells.
\paragraph{Phase noise.}
Shot-to-shot fluctuations of the superlattice relative phase result in random offsets within the double-well and lead to dephasing.
The phase fluctuations measured in \figref{fig:figS1}b translate to offset fluctuations of about \SI{12}{\percent} of the intra-well coupling $t$.
Nevertheless, we find numerically that the induced dephasing remains limited on the timescale of the experiment, with more than \SI{85}{\percent} of the coherence maintained after 50 oscillations.
Over the duration of data acquisition (about 4 hours), we expect the effects of phase drifts to be on the order of shot-to-shot fluctuations, according to the inset of \figref{fig:figS1}b.
\paragraph{Amplitude inhomogeneities.}
The finite extent of the lattice potential results in inhomogeneous double-well coupling strengths across the sample, with larger couplings (and thus larger frequencies) at the edge of the system.
The single-site resolution of our apparatus allows to spatially resolve and quantify such an effect, as shown in \figref{fig:fig2}b.
The data presented in \figref{fig:fig2}a corresponds to a subset of double-wells, such that inhomogeneous coupling effects remain limited in the efficiency of the oscillations.
In particular, we measure a dispersion of double-well couplings among the selected double-well of about \SI{0.5}{\percent}, leading to a Gaussian decoherence with a $\ee^{-1/2}$ lifetime of about 34 oscillations.
\paragraph{Coupling to neighboring double wells.}
The finite depths of our lattices results in a small, but finite tunnel coupling between double-wells, directly affecting the contrast of the oscillation.
We numerically evaluate the effects of such a process by modeling a two-level system (the double-well) coupled to a bath \cite{dalibard:1992}.
As our experimental post-selection ensures that only double-wells containing exactly one detected atom are considered in the evaluation, the incoherent coupling between the system and the bath is modeled as an exchange of atoms due to inter-DW tunneling events.
Specifically, each ``exchange'' event is modeled by a random projective measurement of the atom into one of the two wells.
The population of the neighboring double-wells, moreover, potentially forbids tunneling events associated to atoms escaping.
Such a ``blockade'' effect is approximated in the numerics by halving the escape rate, assuming the neighboring sites are only occupied half of the time, and neglecting the effect due to spins and interactions.
The resulting time evolution shows that coupling between neighboring double-well has the most dramatic effect on the coherence time of the oscillations.
The associated timescales agree well with the measured evolution (\figref{fig:fig2}c), with an exponential decay of $\tau_\mathrm{d} \approx 30 \times 2\pi/\omega$.
\paragraph{Detection fidelity corrections.}
As mentioned in the main text, our postselection analysis allows to take into account most of the state preparation and measurement errors.
Small corrections arise due to our imaging fidelity, during which each atom can be lost or can tunnel to another lattice site, with probability $p_{\mathrm{L}}$ and $p_{\mathrm{T}}$ respectively, independent of the lattice configuration before imaging.

We evaluate here the probability $P_{M_1}(W)$ that the measured configuration, in a given double-well, is wrong ($W$), knowing that a single atom was measured ($M_1$).
We consider here that there can be at most $1$ atom per lattice site.
Keeping terms up to $2^{\mathrm{nd}}$ order in $p_{\mathrm{L}}, p_{\mathrm{T}}$, we find
\begin{widetext}
\begin{multline}
    P_{M_1}(W) = \frac{P(W \cap M_1)}{P(M_1)} \\ = \frac{P_{00}\cdot 2p_\mathrm{T}(1-p_\mathrm{T}) + P_{11}\cdot 2(p_\mathrm{L} + p_\mathrm{T})(1-p_\mathrm{L}-2p_\mathrm{T}) + (P_{01} + P_{10})\cdot p_\mathrm{T}(p_\mathrm{L} + p_\mathrm{T})}{P_{00}\cdot 2p_\mathrm{T}(1-p_\mathrm{T}) + P_{11}\cdot 2(p_\mathrm{L} + p_\mathrm{T})(1-p_\mathrm{L}-2p_\mathrm{T}) + (P_{01} + P_{10})[1 - (p_\mathrm{L} + p_\mathrm{T})(2 - 3p_\mathrm{T}) - 2p_\mathrm{T}^2]},
\end{multline}
\end{widetext}
with $P_{ab}$, ($a, b = 0, 1$) the probability that the double-well configuration is $(a, b)$.
Experimentally, we evaluate $p_\mathrm{T} \approx \num{0.02}$ and $p_\mathrm{L} \approx \num{0.01}$ in the region on which the analysis is performed (see \figref{fig:fig2}b).
In the case where the expected configuration is $(0, 1)$ --- for instance after a $\pi$-pulse starting from $(1, 0)$ --- and taking $P_{01} = P_{11} = P_{00} = \num{5e-3}$ based on the preparation fidelity evaluated from the imbalance measurements in \figref{fig:fig1}d, we find $P_{M_1}(W) = \num{1.8e-3}$.

\subsection{Hubbard parameters estimation}
\label{sec:sechub}

The Hubbard parameters, \emph{i.e.} the tunneling amplitudes $t_{ij}$, the on-site interaction strength $U$, and the staggering energy offsets $\Delta_i$ are estimated from the lattice depths $V_x, V_{x,\mathrm{L}}$ and $V_y$, and the strength of the contact interaction which is determined by the $s$-wave scattering length being set by the strength of the magnetic field.
The lattice depths are calibrated using parametric heating spectroscopy to a precision of about \SI{5}{\percent}, which is the main source of uncertainty in the calculation of the Hubbard parameters.
Then, the uncertainty associated with a given Hubbard parameter is considered to be the standard deviation of the distribution of the parameter, resulting from the lattice depths distributions. These distributions are assumed to follow a Gaussian law with a \SI{5}{\percent} standard deviation.

In the presence of a superlattice, care has to be taken in the choice of the phase of the Wannier function, as this influences the calculation of the Hubbard parameters.
Here, the calculation is performed in the basis of maximally localized Wannier states \cite{marzari:2012, modugno:2012}.
Numerically, these states are computed as band-projected eigenstates of the position operator \cite{bissbort:2012}.

\subsection{Effective Hubbard parameters in staggered configurations}

In the presence of strong staggering, $\Delta \gg t$, NN tunneling is completely suppressed, but NNN remain coupled by two distinct processes: direct coupling, of amplitude $t'$, and perturbative coupling, of amplitude $-t^2/\Delta$.
Here, the NN tunneling amplitude $t$, which corresponds to the overlap of NN Wannier functions, remains non-zero and depends on the energy offset $\Delta$.
Referring to \emph{suppressed} NN tunneling thus refers to the absence of population transfer between nearest neighbors.

Care has to be taken in the consideration of the sign of the tunneling amplitudes.
NN and NNN tunneling amplitudes, in particular, have opposite signs.
In the main text, we consider that $t > 0$, such that the Hubbard Hamiltonian is correctly represented by \Eqref{eq:eq2}.
Keeping the same convention, we consider here that $t' < 0$.
Nevertheless, it is important to point out that, in the perturbative expansion, the \emph{true} coupling, \emph{i.e.} the value of $\braket{i|\hat{H}^\mathrm{eff}|i+2}$, with $\hat{H}^\mathrm{eff}$ the effective Hamiltonian that couples NNN sites, takes the sign of $-\Delta$.
The perturbative and direct couplings, as such, have \emph{opposite} signs if $\Delta > 0$ --- this is the case in \figref{fig:fig3}b, in which the dynamics takes place on the deepest sites --- and have the \emph{same} sign if $\Delta < 0$ --- as in \figref{fig:fig3}c, where the delocalization occurs on the upper sites.

The effective NNN coupling thus becomes
\begin{equation}
    \tilde{t} = \frac{t^2}{\Delta} + t' \quad \text{with $t' < 0$},
\end{equation}
such that, following our convention, the effective Hamiltonian of a single particle in 1D, writes $\hat{H}^\mathrm{eff}~=~-\tilde{t}\sum_{i}\ket{i}\bra{i+2}~+~\mathrm{h.c.}$.
The validity of this effective description of our system is detailed in \secref{sec:secqw}.

A similar line of thought can be carried out for the dynamics of a repulsively-bound pair.
As mentioned in the main text, the motion of the pair on a sublattice is even more strongly suppressed, as it requires both particles to tunnel twice, thus an amplitude scaling as $t^4/\Delta^3$.
The initial breaking of the pair, however, relies on the virtual population of a state in which only one atom has tunneled and shifted in energy by $-\Delta$.
The effective coupling term inducing the pair breaking process thus takes an amplitude $t^2/\Delta$ as well.

\subsection{Quantum walks}
\label{sec:secqw}

In this section, we discuss the validity of the effective description of our quantum walks in staggered lattices and discuss the effects of lattice imperfections.
~ \\
\paragraph{Single-particle quantum walks in standard lattices.}
A single-particle quantum walk in a uniform, standard lattice is described by the single-particle Hamiltonian $\hat{H} = -t\sum_{x}\left(\ket{x}\bra{x + a} + \mathrm{h.c.}\right)$, where $a$ is the lattice constant and $\ket{x}$ the state in which the particle is localised at position $x$.
The dynamics associated to an initial state $\ket{\psi(\tau = 0)} = \ket{0}$ leads to
\begin{equation}
\label{eq:eqS3}
p_x(\tau) = \left|\braket{x | \psi(\tau)}\right|^2 = \left|J_{x}(2t\tau/\hbar)\right|^2,
\end{equation}
where $p_x(\tau)$ the population in site $x$ at time $\tau$ and with $J_x(z)$ the Bessel integral of the first kind.

\begin{figure}[!ht]
\centering
\includegraphics[scale = 1]{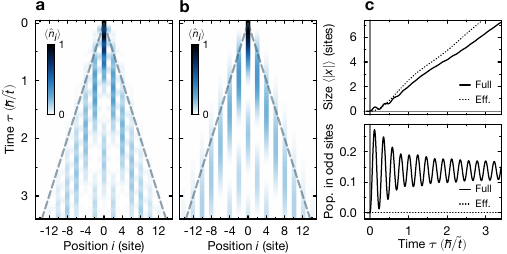}
\caption{
\textbf{Comparison of the full and effective Hamiltonians} \textbf{a}, Full Hamiltonian dynamics and \textbf{b}, effective Hamiltonian dynamics in the case of the Hubbard parameters of \figref{fig:fig3}b.
\textbf{c}, Comparison of the size of the distribution (see main text for the definition) and population in odd sites of the lattices --- the latter are equal to $0$ in the case of the effective Hamiltonian, which does not couple odd and even sublattices.
}
\label{fig:figS3}
\end{figure}

\paragraph{Single-particle in staggered lattices.}
We compare in \figref{fig:figS3} the single-particle dynamics in a staggered configuration calculated using the full Hamiltonian (see \secref{sec:secnum}) and using the effective description based on perturbative couplings and \Eqref{eq:eqS3}.
The parameters used in the simulation are the same as in the experiment (\figref{fig:fig3}b).
The effective description manages to grasp the main properties of the evolution.
Nevertheless, slight differences are visible, in particular in the population of odd sites of the lattice --- they are completely decoupled in the effective description ---, and in the size of the distribution, that increases by a rate being slower by about \SI{15}{\percent}.
\paragraph{Effect of lattice imperfections.}
The finite extent of the lattice beams yields a position-dependent lattice depth of the form $V_x(x) = V_x\ee^{-2x^2/w_x^2}$, with $w_x \approx \SI{120}{\um}$ the $1/\ee^2$ radius of the beam.
Our short lattice, which is the only one used for the experiment presented in \figref{fig:fig3}a, is repulsive for the Li atoms, such that the finite extent of our lattice beams translates into a position-dependent shift of the ground band energy of the lattice.
At $V_x = \SI{10}{\Erx}$, the curvature expected from such a dependence of the ground band energy with position is about $h\times\SI{1.8}{\Hz}/\mathrm{site}^2$.
In the quantum walk experiments presented in \figref{fig:fig3}, the depth of the long lattice does not exceed a few \si{\Erxl}, while the $1/\ee^2$ radius of the beam is on the order of \SI{300}{\um}.
The expected curvature induced by the long lattice is about $h\times\SI{0.18}{\Hz}/\mathrm{site}^2$ for $V_{x,\mathrm{L}} = \SI{5}{\Erxl}$

\begin{figure}[!ht]
\centering
\includegraphics[scale = 1]{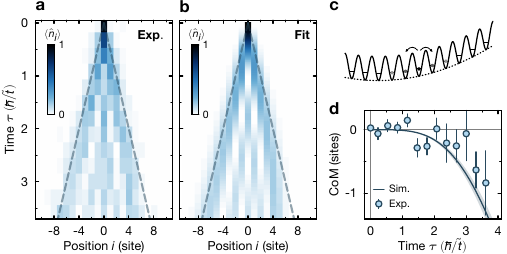}
\caption{
\textbf{Effects of large-scale inhomogeneities}
\textbf{a}, Experimental quantum walk dynamics, as shown in \figref{fig:fig3} in the main text.
\textbf{b}, Result of a fitting procedure, with a model taking into account large scale inhomogeneities of the potential.
\textbf{c}, Cartoon representation of the model, in which the lattice potential has a harmonic envelope and the initial site is chosen to be off-center.
\textbf{d}, Comparison of the center-of-mass (CoM) shift of the distribution from the experimental data (points) and from the result of the fitting procedure (solid line). 
}
\label{fig:figS4}
\end{figure}

The asymmetry of the distribution of populations in the long-time quantum walk dynamics cannot be explained solely by curvature of the lattice, but is rather a consequence of both curvature and ``offset'' of the initial lattice site with respect to the lowest energy site.
We show in \figref{fig:figS4} a comparison of the experimental data with a model taking into account large scale inhomogeneities of the underlying trapping potential, implemented by setting the on-site offset on site $i$ to be $\Delta_i = \gamma(i - i_0)^2$.
The quantum walk dynamics obtained from the model are fitted to the data, leaving the NN tunneling amplitude, the curvature and the offset as free parameters.
The model manages to reproduce the shift of the center-of-mass of the distribution, with fitted parameters $t = 2\pi\times\SI{110(2)}{\Hz}$, $\gamma = \SI{4.7(8)}{\Hz}/\mathrm{site}^2$ and $i_0 = \num{6(1)}$ sites.

\subsection{Numerical simulations}
\label{sec:secnum}

Every numerical result presented in the main text is obtained using exact diagonalization (ED) methods.
~ \\
\paragraph{Imbalance.}
The solid lines in \figref{fig:fig1}d,e correspond to the imbalance of the ground state $\psi_0$ of the Fermi-Hubbard Hamiltonian $\hat{H}$ on a double well, using the Hubbard parameters $t_x^{(1)}$ and $U$ as shown in \tabref{tab:tab1}.
We find that the finite duration of the freezing ramp that we use to measure the atomic distribution slightly impacts the measured result of these measurements (dashed lines in \figref{fig:fig1}d,e).
The effect of the ramps are taken into account by integrating the time-dependent Schr\"odinger equation $\ii\hbar\partial_\tau\ket{\psi} = \hat{H}(\tau)\ket{\psi}$, where the time dependence of the Hubbard parameters is parametrized in $\hat{H}(\tau)$ following the ramp used in the experiment.
~\\
\paragraph{Quantum walks.}
For the quantum walks, the system considered in the numerics is a 1D chain of size $L = 27$ sites, populated with one particle (for single-particle walks) or two interacting particles (for the correlated walks of \figref{fig:fig3}c-e).
The initial state consists in populating the central site $i=0$ only.
The evolution is then simulated \emph{via} unitary evolution under the Fermi-Hubbard Hamiltonian \eqref{eq:eq2}, in which $\Delta_i = \Delta$ ($\Delta_i = 0$) for $i$ odd (even).
~ \\
\paragraph{Ferromagnetic ladders.}

The experimental data presented in \figref{fig:fig4} is compared to numerics obtained from exact finite temperature ED of Fermi-Hubbard ladders on a $2\times4$ system.
The ED is facilitated using translational invariance along $y$, enforced by periodic boundary conditions.

\begin{figure}[!t]
\centering
\includegraphics[scale = 1]{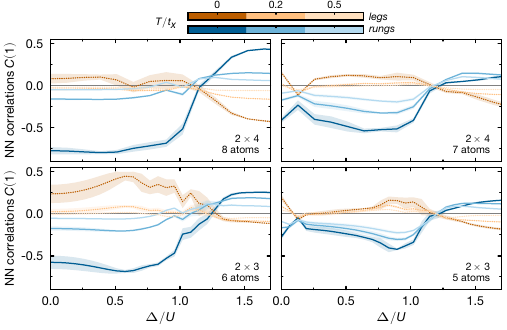}
\caption{
\textbf{Exact diagonalization of Fermi-Hubbard ladders}.
NN spin correlations at different temperatures obtained from exact diagonalization of Fermi-Hubbard ladders (see text).
Top row: $2 \times 4$ system, bottom row: $2 \times 3$ system.
Left column: half-filling, right column: single dopant.
Rung correlations are given in the blue solid curves, leg correlations are given in the dotted brown curves.
}
\label{fig:figS5}
\end{figure}

\begin{figure*}[!t]
\centering
\includegraphics[scale=1]{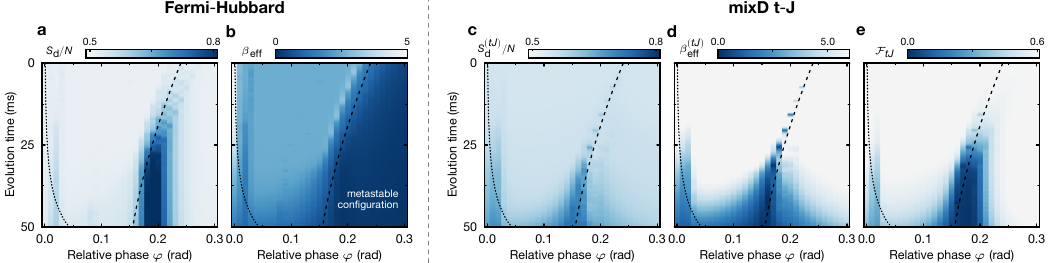}
\caption{
\textbf{Numerical simulations: preparing effective mixD $t-J$ systems}.
\textbf{a}, Diagonal entropy per particle $S_\mathrm{d}/N$ (in units of $k_\mathrm{B}$) and \textbf{b}, effective inverse temperature $\beta_\mathrm{eff}$ (in units of $t_x^{-1}$) as a function of time and relative phase of the superlattice.
\textbf{c}, Diagonal entropy per particle and \textbf{d} effective temperature of the projected state onto the $t-J$ basis (see text).
\textbf{e}, Fidelity of the state to an equilibrium state of the mixD $t-J$ model (see text).
In all plots, the dashed line indicates the resonance condition $\Delta = U$ along the evolution, and the dotted line marks $\Delta = 4t_x$ below which tunneling is not fully suppressed.
}
\label{fig:figS6}
\end{figure*}

Time evolution is simulated by integrating the time-dependent Schr\"odinger equation $\ii\hbar\partial_\tau\hat{\rho} = [\hat{H}(\tau), \hat{\rho}]$, with $\hat\rho$ representing the thermal state of the system.
The initial state, in particular, is chosen to be the thermal state associated to the decoupled ladder used as an intermediate state in the preparation (\figref{fig:fig4}a-(ii)), \emph{i.e.} $\hat\rho(\tau=0)~=~(Z)^{-1}\sum_n\ee^{-\beta E_n}\ket{\phi_n}\bra{\phi_n}$, with $\beta = (k_\mathrm{B}T)^{-1}$ the inverse temperature, $Z = \sum_n\ee^{-\beta E_n}$ the partition function and $E_n, \ket{\phi_n}$ the eignenenergies and eigenstates associated to the decoupled ladder Hamiltonian, $\hat{H}^{\mathrm{(ii)}}\ket{\phi_n} = E_n\ket{\phi_n}$.
The time-dependence of $\hat{H}$ is implemented using the ramps programmed in the experimental sequence (see \appref{sec:appC}), starting from step (iii), \emph{i.e.} a metastable situation in which one leg is higher in energy while there are equal populations in both legs.

Observables of interest --- occupation, spin-correlations --- are evaluated along the time evolution, and the mean and standard deviation of the values over the last \SI{5}{\percent} of the ramp are identified as the final value and uncertainty, to be compared to the experimental data.
Finite size offsets are treated by subtracting the results obtained for $\beta = 0$.

Our numerical simulations show that spin correlations have a substantial dependence on doping and system size (see \figref{fig:figS5}).
We do not expect to obtain a quantitative agreement between the ED and the experimental result, the latter being obtained on a much larger system.
Qualitatively, however, the ED manages to reproduce the change of sign of the rung correlations for $\Delta > U$, as well as the boost of the rung correlations for intermediate values $0 < \Delta < U$, irrespective of the system size and doping.

In order to assess the quality of our ramp, we extract from the numerical simulations the diagonal entropy $S_\mathrm{d}(\tau) = -k_\mathrm{B}\sum_np_n(\tau)\ln[p_n(\tau)]$, with $p_n(\tau) = \braket{\phi_n(\tau)|\hat\rho(\tau)|\phi_n(\tau)}$ the population in eigenstate $\ket{\phi_n(\tau)}$ at time $\tau$.
Quite generally, the diagonal entropy quantifies the adiabaticity of dynamics in closed quantum systems \cite{polkovnikov:2011a, polkovnikov:2011}.
Here, we focus on a system initialized at $\beta t_x = 5$, consistent with the experimental situation $T/t_x \approx 0.2$.
We define an effective inverse temperature $\beta_{\mathrm{eff}}$, defined as $S_\mathrm{eq}(\beta_\mathrm{eff}) \equiv S_\mathrm{d}$, with $S_\mathrm{eq}(\beta)$ the equilibrium entropy at inverse temperature $\beta$.
This analysis is performed on a system of size $2\times 4$ with a single dopant.

We show in \figref{fig:figS6}a $S_\mathrm{d}$ as a function of superlattice phase and evolution time.
We do not observe a significant increase of $S_\mathrm{d}$ during the evolution, except close to the doublon resonance (dashed lines), as expected from the coupling to additional states in the system, and at low tilts $\Delta \lesssim 4t_x$ (dotted lines) where tunneling between legs is not fully suppressed.
As such, the effective inverse temperature $\beta_{\mathrm{eff}}$ (\figref{fig:figS6}b) does not decrease significantly below resonance ($\Delta < U$).
Above resonance ($\Delta > U$), however, the system is in a metastable state, for which an effective temperature cannot be properly defined.

Away from resonance, the system is expected to be well described by the $t-J$ Hamiltonian in mixed dimensions (mixD)
\begin{align}
\hat{H}_{tJ} =\ & \hat{\mathcal{P}}^\dag\sum_{\braket{ij},\sigma}\left[-t_{ij}\hat{c}^\dag_{i,\sigma}\hat{c}_{j,\sigma} + \mathrm{h.c.}\right]\hat{\mathcal{P}} \notag \\ & + \sum_{\braket{ij}}J_{ij}\left(\hat{\bm{S}}_i\cdot\hat{\bm{S}}_j - \frac{\hat{n}_i\hat{n}_j}{4}\right),
\end{align}
with $t_{ij} = 0$ if $i,j$ are NN along the rungs ($x$) as a result of the tilt.
Here, $\hat{\mathcal{P}}$ projects out the doubly occupied states, $J_{ij}~=~4t_y^2/U$ if $i,j$ are NN along the legs of the ladder ($y$), and $J_{ij}~=~J(\Delta)$ (see \Eqref{eq:eq3} in the main text) for $i,j$ NN along the rungs ($x$).
In order to compare our system to the effective $t-J$ model, we consider the projected density matrix $\hat{\rho}_{tJ}(\tau) = \hat{\mathcal{P}}\hat{\rho}(\tau)\hat{\mathcal{P}}^\dag$, and conduct a similar analysis to the one described above.
In particular, we extract a diagonal entropy $S_\mathrm{d}^{(tJ)}$ and an effective inverse temperature $\beta_{\mathrm{eff}}^{(tJ)}$ by comparing to the equilibrium entropy of the associated mixD $t-J$ model.
The results shown in \figref{fig:figS6}c,d indicate that, in the regime where motion along the rungs is expected to be suppressed, \emph{i.e.} for $t_x < \Delta < U$ and for $\Delta > U$, the increase of entropy and effective temperature remains limited. We estimate how well the mixD $t-J$ model is realized by evaluating the fidelity $\mathcal{F}_{tJ}(\tau) = \mathrm{Tr}[(\hat{\rho}_{tJ}(\tau)\hat{\rho}_{tJ}^{(\mathrm{eq})})^{1/2}]^2$, with $\hat\rho_{tJ}^{(\mathrm{eq})}$ the equilibrium density matrix associated to the mixD $t-J$ model at inverse temperature $\beta_{\mathrm{eff}}^{(tJ)}$.
The results are shown in \figref{fig:figS6}e.
In the regime where motion is suppressed, the fidelity does not change with time and remains close to $\mathcal{F}_{tJ} \approx 0.6$. 
\end{document}